\documentclass[manuscript]{aastex}

\def\simge{\mathrel{%
    \rlap{\raise 0.511ex \hbox{$>$}}{\lower 0.511ex \hbox{$\sim$}}}}
\def\simle{\mathrel{
    \rlap{\raise 0.511ex \hbox{$<$}}{\lower 0.511ex \hbox{$\sim$}}}}

\slugcomment{submitted to Ap. J. Lett.}
\shorttitle{Thermometric Soots}
\shortauthors{Zahnle et al.}

\begin{document}
\title{Thermometric Soots on Warm Jupiters}

\author{K.\ Zahnle}
\affil{NASA Ames Research Center, Moffett Field, CA 94035}
\email{Kevin.J.Zahnle@NASA.gov}

\author{M.\ S.\ Marley}
\affil{NASA Ames Research Center, Moffett Field, CA 94035}
\email{Mark.S.Marley@NASA.gov}

\and
\author{J.\ J.\ Fortney}
\affil{Department of Astronomy and Astrophysics, University of California - Santa Cruz}

\DeclareGraphicsRule{.tif}{png}{.png}{`convert #1 `dirname #1`/
`basename #1 .tif`.png}

\begin{abstract}

We use a 1D thermochemical and photochemical kinetics model to predict %
the disequilibrium stratospheric chemistries of warm and hot Jupiters ($800\!<\!T\!<\!1200$ K).
 Thermal chemistry and vertical mixing are generally more important than photochemistry.
At 1200 K, methane is oxidized to CO and CO$_2$ by OH radicals from thermal decomposition of water.
At $T<1000$ K, methane is reactive but stable enough to reach the stratosphere, 
while water is stable enough that OH levels are suppressed by reaction with H$_2$. 
These trends raise the effective C/O ratio in the reacting gases above unity.
Reduced products such as ethylene, acetylene, and hydrogen cyanide become abundant;
further polymerization should lead to formation of PAHs (Poly-Aromatic Hydrocarbons) and soots.
Parallel shifts are seen in the sulfur chemistry, with CS and CS$_2$ displacing S$_2$ and HS as the interesting disequilibrium products. 
Although lower temperature is a leading factor favoring hydrocarbons, 
higher rates of vertical mixing, lower metallicities, and lower incident UV radiation also favor organic synthesis.
Acetylene (the first step toward PAHs) formation is especially favored by high eddy diffusion coefficients $K_{zz}>10^{10}$ cm$^2$/s. 
In most cases planetary compositions inferred from transit observations
will differ markedly from those inferred from reflected or emitted light from the same planet.
The peculiar properties of HD 189733b compared to other hot Jupiters --- a broadband blue haze, little sign of Na or K, and hints of low metallicity --- can be
explained by an organic haze if the planet is cool enough.
Whether this interpretation applies to HD 189733b itself, many organic-rich
warm Jupiters are sure to be discovered in the near future.  

\end{abstract}


\section{Introduction}

The most easily studied transiting hot Jupiters are HD 209458b and HD189733b (Fortney et al 2009).
HD 209458b is typical of known hot Jupiters (Burrows et al.\ 2008).
Transit observations show a visible spectrum dominated by the 590 nm sodium resonance lines (Charbonneau et al 2002),
the visibility of which implies a clean, haze-free atmosphere (Sing et al 2008).
%
HD 189733b is atypical, although its details are controversial (Fortney et al 2009).
In visible light HD 189733b presents a featureless blue haze with little sign of the expected Na and K lines (Pont et al 2008).
Lecavalier des Etangs et al (2008) suggest that these lines are obscured by enstatite grains mixed up to microbar levels.
Swain et al (2008, 2009) fit transit data between 1.5 and 2.5 $\mu$m with CO, CO$_2$, H$_2$O, and CH$_4$,
all with distinctly subsolar abundances corresponding to a metallicity of $[{\rm M}/{\rm H}]\approx -0.5$.
They suggested that adding C$_2$H$_2$, C$_2$H$_4$, or possibly NH$_3$, would improve their model's fit.
On the other hand Fortney et al (2009) fit D{\'e}sert et al's (2009) dayside spectrum at 4.5 $\mu$m with supersolar CO and CO$_2$ corresponding to $0.5<[{\rm M}/{\rm H}]<1.5$. 

An organic haze offers another way to explain the appearance of HD 189733b.
Here we use a photochemical model 
to address the threshold for organic haze formation in hot or warm Jupiters.
In Zahnle et al (2009) we noted that temperatures below 1200 K 
change the kinds of chemistry that take place.
In hot atmospheres methane is efficiently oxidized to CO.
Cooler atmospheres are functionally more reduced because H$_2$O is more stable: 
organic molecules are favored over CO, PAHs (polycyclic aromatic hydrocarbons) and soots (disorganized agglomerations of bigger PAHs) might form and precipitate.  
The PAHs and soots would be the source of the haze.

\section{Previous Models}

The possibility that organic hazes might be important in irradiated brown dwarfs was first raised by Griffith et al (1998).
However, equilibrium chemistry does not predict the presence of organic hazes.
Nor did previous photochemical models (Liang et al 2003, Liang et al 2004).
Liang et al (2004) showed that simple hydrocarbons would not condense
to form photochemical smogs in hot solar composition atmospheres.
They concluded that photochemical smogs would not be present in the hottest hot Jupiter atmospheres that they addressed,
but they did not consider cooler atmospheres where hydrocarbons would be more stable.

Zahnle et al (2009) developed a new 1D atmospheric chemistry model
that addresses photochemical and thermochemical disequilibrium in hot Jupiter atmospheres. 
The focus of our first paper was on sulfur chemistry at temperatures between 1200 and 2000 K. 
Sulfur is a relatively abundant volatile element that equilibrium calculations suggest should be
present as H$_2$S (Visscher et al 2006).
Our disequilibrium calculations suggest that sulfur should be present at observable altitudes as S$_2$ or HS.
Sulfur is important to the chemistry because H$_2$S is the most
reactive primary molecule in solar composition atmospheres at temperatures between 800 K and 2000 K.
Its thermal decomposition is the biggest source of free radicals and H in particular.
We found that HS and S$_2$ should be abundant at the millibar level,
and we suggested that most incident light from 250 nm to 450 nm would be absorbed by stratospheric S$_2$ and HS.
The resultant stratospheric heating can be considerable in planets that are not hot enough for TiO and VO to evaporate 
in significant quantities.

In passing we also confirmed an earlier suggestion (Fegley and Lodders 2002, Visscher et al 2006) that CO$_2$ should be a very good indicator of metallicity. 
We showed that this result applies when kinetics are included and
that the CO$_2$ abundance and the CO$_2$/CO ratio are insensitive to $T$ in the range $1200<T<2000$ K. 

\section{Model}

We use a standard 1D kinetics code to simulate atmospheric chemistry of hot irradiated giant planets.
Volume mixing ratios $f_i$ of species $i$ are obtained by solving continuity
\begin{equation}
\label{eq_one}
N{\partial f_i \over \partial t} = P_i - L_i N f_i - {\partial \phi_i \over \partial z} 
\end{equation}
\noindent and force (flux)
\begin{equation}
\label{eq_two}
\phi_i = b_{ia} f_i \left( {m_a g\over kT} - {m_i g\over kT}\right) - \left( b_{ia} + K_{zz}N\right) {\partial f_i \over \partial z}
\end{equation}
\noindent equations.
In these equations $N$ is the total number density (cm$^{-3}$); $P_i - L_i N f_i$ represent chemical production and loss terms, respectively; $\phi_i$ is the upward flux; $b_{ia}$, the binary diffusion coefficient between $i$ and the background atmosphere $a$, describes true molecular diffusion; $m_a$ and $m_i$ are the molecular masses of $a$ and $i$; and $K_{zz}$, the eddy diffusion coefficient, parameterizes vertical mixing as diffusion.
We have implemented molecular diffusion through H$_2$
by setting $b_{ia} = 6\times 10^{19}\left(T/1400\right)^{0.75}$ cm$^{-1}$s$^{-1}$ --- appropriate for CO --- for all the heavy species.
This is a reasonable choice for present purposes.
The physical meaning of $b_{ia}$ is the ratio of the relative thermal velocities of two species to their collision cross section. 
In the present circumstances, the relative thermal velocity is effectively that of H$_2$, so the only important source of variation in $b_{ia}$ 
is in the different diameters of the molecules.

In practice Eqns \ref{eq_one} and \ref{eq_two} are solved as a system of second order partial differential equations for $f_i(z,t)$.
Steady state solutions are found by integrating the equations through time (typically $10^9$ years) 
using a fully implicit backward-difference method.
Some aspects of the code are briefly described in Zahnle et al (2009).
Other aspects are more completely described as applied to ancient Earth (Zahnle et al 2007)
and Mars (Zahnle et al 2008). Here we fully describe the chemical system.

In this study we solve 528 chemical reactions and 33 photolysis reactions for 58 chemical species.
The 58 species are listed in Table 1.
The reactions and references are listed, and where appropriate discussed, in Tables 2 and 3.
In our scheme the hydrocarbon chemistry is truncated at C$_2$H$_m$ (with the exceptions of C$_4$H and C$_4$H$_2$).
This means that polymerization beyond C$_2$H$_m$ is not included. 
Thus when conditions favor polymerization carbon pools in C$_2$H$_m$, because longer
carbon chains are not allowed.

New to this model are several hydrocarbon species added since Zahnle et al (2009).
Methanol (CH$_3$OH) and the radicals CH$_3$O and H$_2$COH allow us to fully describe the known
hydrogenation pathways from CO to CH$_4$.  These pathways need to be included because equilibrium chemistry 
predicts that CO should convert to CH$_4$ under warm Jupiter conditions.  In practice these reactions are quite unimportant and
CO is effectively indestructible under any conditions encountered in this study. 
Given that CO does not hydrogenate by gas phase reactions to any significant extent, we see no call at this time to include speculative pathways through the extremely unstable N$_2$H radical that might allow hydrogenation of N$_2$ to NH$_3$.

In the original preprint edition of this paper we did not include addition reactions of hydrocarbons with OH, such as ${\rm C}_2{\rm H}_4 + {\rm OH} + {\rm M} \rightarrow {\rm adduct} + {\rm M}$. 
The precise nature of the adduct is generally unknown, nor is it known in general how the adduct reacts. 
But as discussed above, the C-O bond created cannot be easily undone in the atmosphere.
In the preprint we considered only the competing H abstraction path ${\rm C}_2{\rm H}_4 + {\rm OH} \rightarrow {\rm C}_2{\rm H}_3 + {\rm H}_2{\rm O}$.
This is a reasonable approximation at high temperatures and low pressures where
the abstraction path is favored over the (3-body) addition. 
But by neglecting the reaction path in which the OH directly attacks a carbon atom, we underestimated the importance of oxidation at higher pressures and lower temperatures,
and thus the preprint overstated the stability of organic molecules.

We have since upgraded the chemical model to explicitly include
 the adduct species C$_2$H$_2$OH, C$_2$H$_4$OH, and HCNOH.
Reaction rates for making the adducts are described in the literature (R510, R512, R514).  
What happens to the adducts subsequently is neither simple nor fully known.
For simplicity we have limited the options to reactions with atomic hydrogen, which is by far the most abundant free radical in these simulations,
and we have chosen simple products that do not artificially reduce the number of free radicals (R511, R513, R515).
The specific products for R511 and R515 were picked by analogy to the documented reaction of C$_2$H$_4$OH (R513). 
These improvements in the reaction scheme lead to moderate changes from the original preprint in the direction expected:
in the new calculations the temperature threshold for hydrocarbon polymerization is about 100 K cooler. 

An important simplification in this study, as in its predecessors,
 is that we assume isothermal atmospheres with constant eddy diffusion coefficients. 
This choice facilitates presenting the results of parameter studies, as $T$ and $K_{zz}$ prove to be the most important parameters.
Published temperature-pressure profiles for hot jupiters are themselves
models.  These models depend explicitly on the assumed equilibrium chemical composition of the atmosphere; the absence (or presence) of arbitrary
amounts of hazes or clouds; time of day or zonal circulation; etc.   
This study emphasizes the chemistry, and its intent is to provide the next step in the iteration between chemical and radiative-convective models.
A ``realistic'' temperature profile requires at least two parameters (an effective temperature and a mean infrared opacity at minimum), 
and at least three more (an internal luminosity, visible opacity, and an insolation geometry) for full fidelity.
Application to more realistic temperature profiles will be deferred to future work.

What $K_{zz}(z)$ should be is not well constrained. 
Values ranging from $10^3$ at the top of the troposphere to $10^7$ cm$^2$/s at the top of the stratosphere
seem to be useful for Jupiter.  Because the amplitudes of upward propagating waves grow as $p^{-0.5}$, 
it is often assumed that $K_{zz} \propto p^{-0.5}$. 
The strong solar forcing of hot Jupiters can lead to much more energetic motions than seen on Jupiter itself.
Showman et al (2009) developed 3D models of the circulation in hot Jupiters  HD 189733b and HD 209458b.
They suggested that ``eddy diffusivities at 1 mbar are $\sim 10^{11}$ cm$^2$ sec$^{Ð1}$.''
For simplicity we assume $K_{zz}$.

The background atmosphere is assumed 84\% H$_2$ and 16\% He.
The relative abundances of C, N, O, and S are solar. 
The elements other than H and He are scaled as a group according to metallicity.
Metallicity is varied from $-0.7 \leq [{\rm M}/{\rm H}] \leq 1.7$.
Surface gravity is $g=20$ m/s$^2$. 
Surface gravity is important to optical depth and therefore 
important to what a planet looks like, but it has little effect on the chemistry (Zahnle et al 2009).
Incident UV radiation is 100$\times$ greater than at Earth and the solar zenith angle is $\theta=30^{\circ}$ unless otherwise noted.
When examining the effect of photolysis we vary the UV flux from 1 to 1000 times solar.

The upper boundary is a zero flux lid at $\sim\!1\mu$bar.
For the lower boundary we assume thermochemical equilibrium mixing ratios of 
 H, H$_2$O, CO, CH$_4$, NH$_3$, N$_2$, and H$_2$S  (Lodders and Fegley 2002, Visscher et al 2006).
Other species are assumed to vanish at the lower boundary.  This ensures that photochemical products
flow into the deep atmosphere, where they are presumptively recycled.

The lower boundary pressure was immaterial to our previous study because temperatures were
very high and the relevant chemical reactions very fast.
But at temperatures below 1000 K, the carbon and nitrogen chemistries become sensitive to the choice of lower boundary conditions. 
Here we impose chemical equilibrium at the lower boundary. 
The puzzle is where to put it.
Chemical equilibrium is not actually expected in our model.
Vertical mixing at 1000 K lifts stable molecules like NH$_3$ faster than it reacts and,
even at 100 bars, gas phase chemistry provides no effective means of converting CO to CH$_4$ and N$_2$ to NH$_3$,
 and is only beginning to convert excess H to H$_2$.
In practice it is likely that CO is converted to CH$_4$ and N$_2$ is converted to NH$_3$ on grains,
in particular Fe-Ni grains if present.
Here we consider arbitrary lower boundary pressures of 1 and 100 bars. 


\section{Results}

\begin{figure}[!htb]
 \centering
    \includegraphics[width=1.0\textwidth]{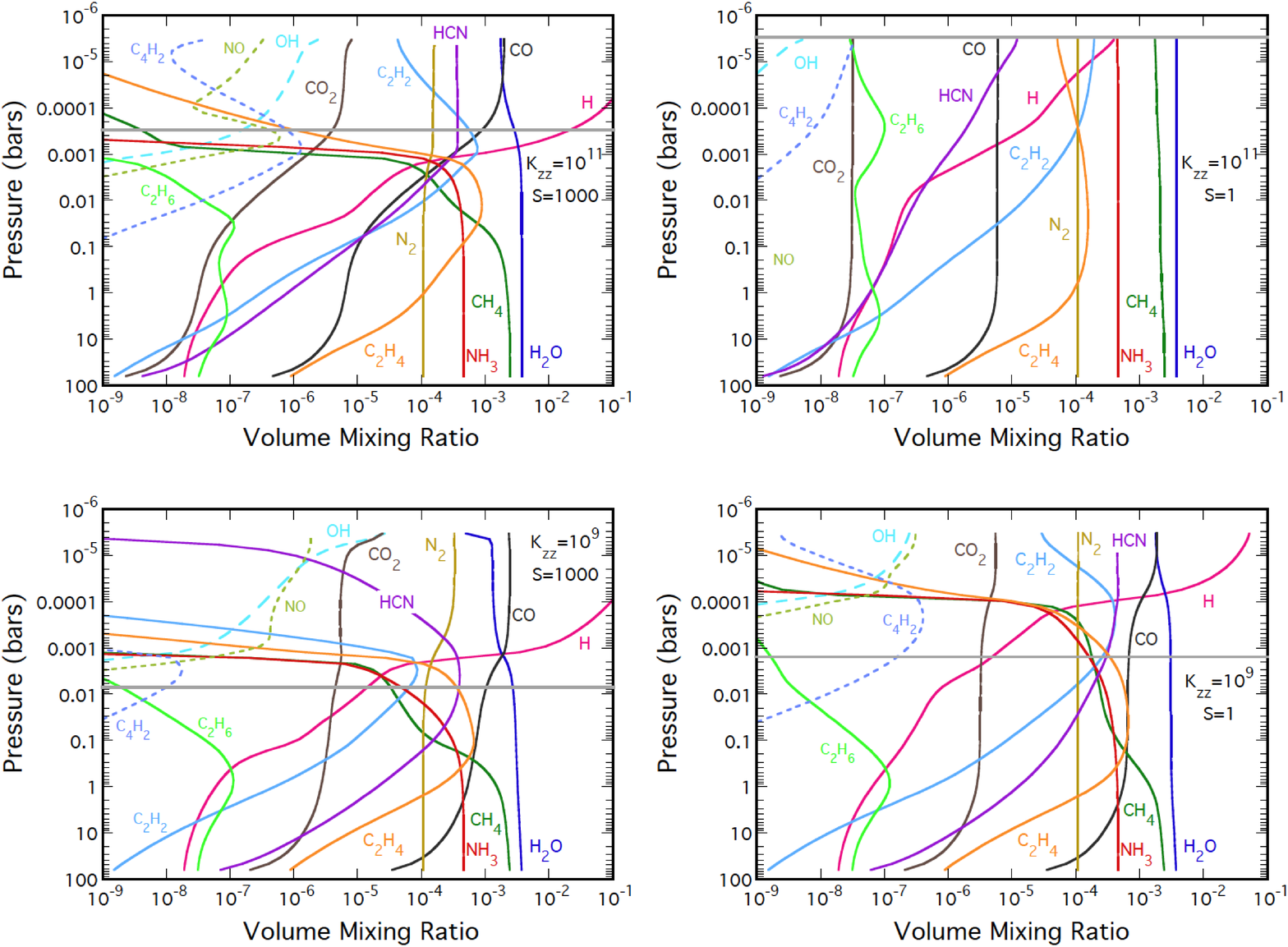} 
 \caption{Vertical profiles of selected H, C, N, O species in isothermal atmospheres 
   at higher and lower levels of UV and EUV irradiation.
   Fixed model parameters are $T=1000$ K, $[{\rm M}/{\rm H}]=0.7$, and $g=20$ m/s$^2$.
   At 1000 K, photolysis is most important when vertical mixing is vigorous ($K_{zz}=10^{11}$ cm$^2$/s).
   Photolysis is less important at $K_{zz}=10^{9}$, and at $K_{zz}=10^{7}$ (not shown) is important only for NH$_3$ and HCN.
   In general, UV is hostile to NH$_3$ and small hydrocarbons other than CH$_4$, and through H$_2$O photolysis creates a more oxidizing environment.
   These effects suppress hydrocarbon formation and relegate hydrocarbon
   formation to a deeper level in the atmosphere.
    Soot's precursors, in particular, are strongly favored by darkness. 
  The horizontal gray bars denote the most favorable altitudes for hydrocarbon formation $p_f$.  For $S=1$ and $K_{zz}=10^{11}$, this would lie above the top of the model.
}
\label{fig:figure_Sun} 
\end{figure}

\begin{figure}[!htb]
 \centering
    \includegraphics[width=1.0\textwidth]{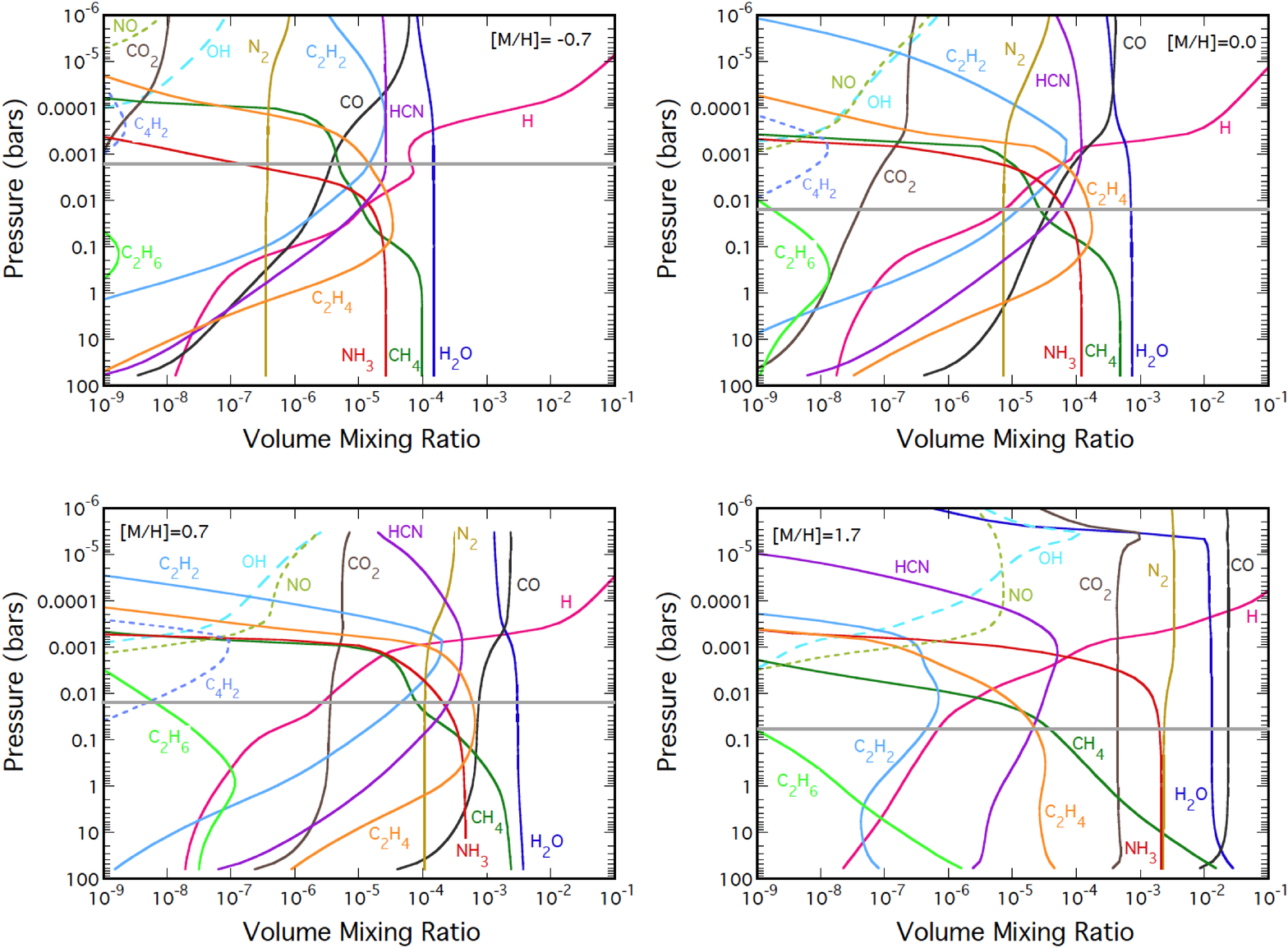} 
  \caption{ Vertical profiles of selected H, C, N, O species in isothermal atmospheres 
   at four metallicities.  Relative abundances of C, N, S, and O are solar.
   Fixed model parameters are $K_{zz} = 10^9$ cm$^2$/s, $T=1000$ K, $S_{uv}=100$, and $g=20$ m/s$^2$.
   The general trend is that higher metallicities result in more strongly oxidized mixtures. 
   CO and especially CO$_2$ are good indicators of metallicity.
   The propensity to form soot appears to be similar for $-0.7 \leq [{\rm M}/{\rm H}] \leq 0.7$, as judged by
   C$_4$H$_2$ abundances, but higher metallicity is unfavorable.   
  The horizontal gray bars denote the most favorable altitudes for hydrocarbon formation $p_f$.
}
\label{fig:figure_Metals}  
\end{figure}

The lower temperatures considered in this study result in a much richer chemistry than we encountered in Zahnle et al (2009).
Temperature, vertical mixing, UV and EUV irradiation, metallicity, and the choice of lower boundary conditions can all be important 
parameters in the 800-1200 K temperature range.
We begin with an investigation of how our results vary with a single parameter when the other parameters are held fixed.
Figures \ref{fig:figure_twelve}--\ref{fig:figure_Metals} are classic spaghetti plots used to show how computed atmospheric compositions vary with altitude.
Figure \ref{fig:figure_twelve} addresses temperature,
Figure \ref{fig:figure_K} addresses vertical mixing,
Figure \ref{fig:figure_Sun} addresses UV irradiation, and
Figure \ref{fig:figure_Metals} addresses metallicity.
Figure \ref{fig:figure_twelve} is also used to illustrate the sensitivity to the lower boundary condition in one particular case,
and Figure \ref{fig:figure_K} is also used to compare computed disequilibrium chemical compositions 
to thermochemical equilibrium compositions at 1000 K.

Spaghetti plots have their place, but it can be hard to see the entr{\'e}e for the noodles.
To describe the bigger picture, we use pie charts that show computed chemical compositions 
at a certain key altitudes across the general parameter survey.
The pies are arrayed in rows of constant $K_{zz}$, according to temperature or metallicity.
We use these to discuss the propensity of some atmospheres to form soots.
The discussion begins with and concentrates on hydrocarbons, but we also provide brief overviews of nitrogen and sulfur chemistry. 

\subsection{Parameters}

\begin{figure}[!htb]
 \centering
    \includegraphics[width=1.0\textwidth]{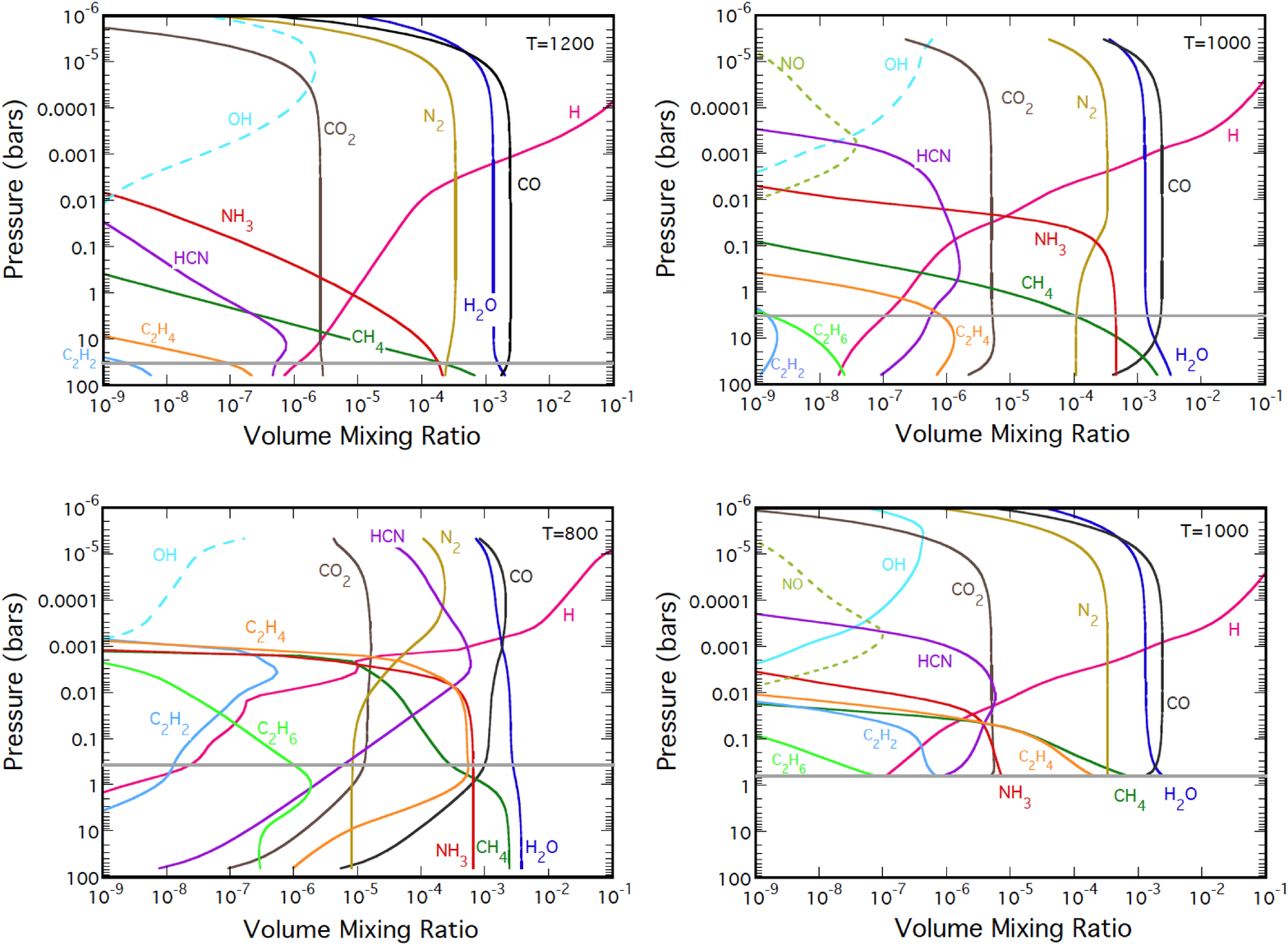} 
   \caption{ Vertical profiles of selected H, C, N, O species in isothermal atmospheres 
   at three temperatures ($T=1200$, 1000, and 800 K).
   Fixed model parameters are $K_{zz} = 10^7$ cm$^2$/s, planetary ($[{\rm M}/{\rm H}]\!=\!0.7$) metallicity, $I=100$, and $g=20$ m/s$^2$.
   The lower boundary condition is thermochemical equilibrium at 100 bars.
   The fourth panel, to be compared to the panel above it,
   raises the thermochemical equilibrium lower boundary to 1 bar.
   At 1200 K methane is oxidized with little buildup of hydrocarbons.
   At 800 K methane's destruction leads to 
   significant disequilibrium production of organic molecules, especially ethylene (C$_2$H$_4$).
  The horizontal gray bars denote the most favorable altitudes for hydrocarbon formation $p_f$ as defined in the text.
  The decreasing abundance of stable species at high altitudes is caused by molecular diffusion.
}
\label{fig:figure_twelve} 
\end{figure}

Figure \ref{fig:figure_twelve} illustrates vertical profiles of representative isothermal atmospheres at 1200, 1000, and 800 K.
These use $K_{zz}=10^7$ cm$^2$/s for vertical mixing and $[{\rm M}/{\rm H}]=0.7$ for metallicity, both comparable to those seen in Jupiter. 
The overall structure of these atmospheres resembles a poorly functioning gas stove, with methane introduced at the
 bottom and flowing upward, where it is either oxidized to CO or made into bigger organic molecules.
At 1200 K, methane is quickly and cleanly oxidized, but at 800 K methane combustion generates mostly ethylene (C$_2$H$_4$) and HCN, to the point where ethylene becomes more abundant than methane.  
This does not mean that C$_2$H$_4$ actually would be more abundant than CH$_4$ (our model truncates polymerization at C$_2$H$_n$),
but it does mean that polymerization is a major methane reaction pathway, and it suggests that hydrocarbons bigger than C$_2$H$_4$ are likely to form.

The fourth panel of Figure \ref{fig:figure_twelve} illustrates the effect of changing the lower boundary pressure. 
Placing the lower boundary at 1 bar puts more methane and less ammonia at 1 bar than there would be if the boundary were at 100 bars.
Ammonia is more abundant in the 100 bar simulations because (i) its equilibrium abundance is higher at higher pressure and (ii) its chemical destruction is kinetically inhibited at 1200 K and below.
Hence the deeper lower boundary puts more NH$_3$ in the stratosphere.
Similar behavior would be seen in CH$_4$ at $T<800$ K.
But at 1000 K, putting the lower boundary at 1 bar
introduces an artificial source of methane that results in a big upward flux that in turn results in more ethylene.
As both the 1 bar and 100 bar lower boundaries are arbitrary in this study  
 it is not useful to take this discussion any further here.
 
The chief factors behind the temperature dependence seen in Figure \ref{fig:figure_twelve} are the high abundances of H and H$_2$ and the different thermal stabilities of methane and water.
Atomic H is more abundant than it would be in equilibrium.
Disequilibrium of H and H$_2$ is closely tied to the disequilibria of H$_2$S, HS, S, and S$_2$.
An important reaction that does not occur at significant rates is the 3-body recombination of H to H$_2$:
$${\rm H} + {\rm H} + {\rm M}  \leftrightarrow {\rm H}_2 + {\rm M} .\eqno{\rm{R}1} $$
The computed H/H$_2$ ratio seen in the kinetics models is much higher than would
be predicted by equilibrium.
High disequilibrium H abundances were also a prominent feature in the models of Liang et al (2003) and Zahnle et al (2009).

Both H$_2$O and CH$_4$ are attacked by H:
$${\rm H} + {\rm CH}_4 \rightarrow {\rm H}_2 + {\rm CH}_3 \eqno{\rm{R}45} $$
$${\rm H} + {\rm H}_2{\rm O} \rightarrow {\rm H}_2 + {\rm OH} . \eqno{\rm{R}6} $$
R6 is significantly endothermic, while R45 is nearly neutral, but
both R45 and R6 have significant activation energies.
Because R6 is endothermic, the temperature barrier is higher, so that water becomes unreactive at a higher temperature than methane,
while the reverse of R6,
$${\rm H}_2 + {\rm OH} \rightarrow {\rm H} + {\rm H}_2{\rm O} . \eqno{\rm{R}5} $$
remains fast at 1000 K.
The net effect of these trends is that, at 1200 K and above, water is very reactive and the gas is generally oxidizing,
while methane reacts so quickly that it fails to reach altitudes significantly above the model's lower boundary.
When the temperature is below 1000 K, water becomes stable enough that OH is scarce,
while CH$_4$ becomes stable enough to be abundant in the atmosphere, yet unstable enough to be very reactive.  
As a consequence the effective C/O ratio in the reacting gases is high and the products are hydrocarbons.

\begin{figure}[!htb]
 \centering
    \includegraphics[width=1.0\textwidth]{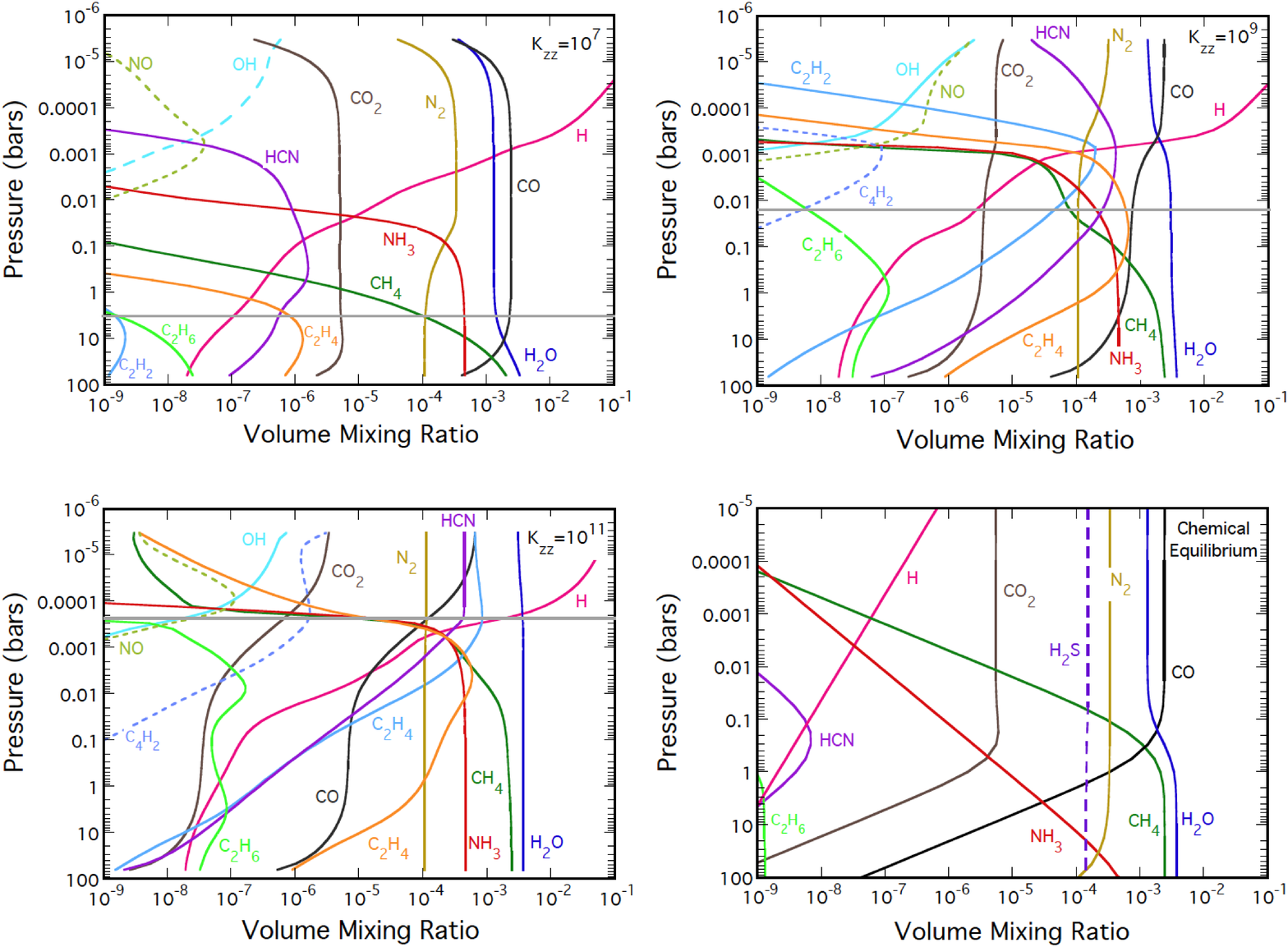} 
  \caption{Vertical profiles of selected species in isothermal 1000 K atmospheres and $[{\rm M}/{\rm H}]=0.7$ as a function of vertical mixing.
   Strong vertical mixing --- $K_{zz} > 10^9$ --- favors disequilibrium production of C$_2$H$_4$, HCN, and especially C$_2$H$_2$ (acetylene, the canonical soot precursor).
  The horizontal gray bars denote the most favorable altitudes for hydrocarbon formation $p_f$ as defined in the text.
  The fourth panel shows thermochemical equilibrium abundances at all altitudes at $T=1000$ and $[{\rm M}/{\rm H}]=0.7$.
  In general, the kinetics calculations predict a much richer variety of species than equilibrium.
}
\label{fig:figure_K} 
\end{figure}

Mismatches between the kinetics model and thermochemical equilibrium at the lower boundary cause flows into or out of the lower atmosphere. 
The most important mismatches stem from kinetic inhibition
of the 3-body reactions that hydrogenate CO to CH$_4$ and N$_2$ to NH$_3$, and the 3-body reactions that reconstitute H$_2$ from H.
The latter results in a high disequilibrium abundance of H.
Very high abundances of H are a prominent feature of all the atmospheres 
in Figures \ref{fig:figure_twelve} and \ref{fig:figure_K}.
The overabundance of H pushes R45 farther to the right than it would be in thermochemical equilibrium.

Figure \ref{fig:figure_K} illustrates the influence of different rates of vertical transport in 1000 K atmospheres.
The overall trend is that faster vertical transport lifts CH$_4$ to higher altitudes, and thus
raises the altitudes where C$_2$H$_2$, C$_2$H$_4$, and HCN form. 
Faster transport increases the efficiency of hydrocarbon formation vis a vis CO formation,
 as also seen in the CO profiles, and it changes the mix of products to favor C$_2$H$_2$ and HCN over C$_2$H$_4$.
 Apparently soot formation is likeliest if $K_{zz}>10^{10}$ cm$^2$/s.
 
The lower right-hand panel of Figure \ref{fig:figure_K} shows what a 1000 K chemical equilibrium atmosphere would look like.
The chemical equilibrium atmosphere can be thought of as having no vertical mixing ($K_{zz}=0$) and no photochemistry ($I=0$),
although it would not be possible to reach equilibrium by setting $K_{zz}=0$ in the kinetics model, because
there are no efficient pathways for making NH$_3$ from N$_2$ or CH$_4$ from CO
(in practice we found it difficult to obtain converged solutions with $K_{zz}<10^5$ cm$^2$/s).
In general, the disequilibrium kinetic atmospheres contain a much richer mix of chemical species.
Product species are generally more abundant in the kinetics models, but so too is ammonia.
On the other hand, at the altitudes where they would be observed, CO and CO$_2$ tend to be present at nearly their
equilibrium abundances.  

Although the disequilibrium chemistry seen in our models is primarily due to thermal chemistry and vertical transport,
photolysis can be important, especially at high altitudes and lower temperatures.
Figure \ref{fig:figure_Sun} compares the effects of photolysis at higher and lower levels of UV and EUV irradiation than used in Figures 
  \ref{fig:figure_twelve} and \ref{fig:figure_K}.
Figure \ref{fig:figure_Sun} shows that UV and EUV photolysis inhibits hydrocarbon formation.
Soot's precursors grow better in the dark.
The consequences of photolysis in warm Jupiter atmospheres are more or less opposite to what is seen on Titan.
In cold atmospheres like Titan's, methane photolysis leads to production of more complicated hydrocarbons.   
But in warm Jupiters, methane is broken up and hydrocarbons are made by thermochemistry.  
Photolysis's role is to free OH radicals from water.
Oxidation by OH inhibits or prevents hydrocarbon polymerization.
All of our atmospheres become more oxidized, and presumably less hazy, at high altitudes due to photolysis of H$_2$O.
In effect, the topmost atmosphere is bleached by photolysis.
The chemocline between the oxidized upper atmosphere and the reduced lower atmosphere is especially pronounced 
in the 800 K atmosphere in Figure \ref{fig:figure_twelve} where the temperature is low enough that thermal decomposition of water is relatively minor.

 Figure \ref{fig:figure_Metals} addresses metallicity.  
Enhanced metallicity favors molecules such as N$_2$ and CS$_2$ that contain multiple elements other than H over molecules such as CH$_4$ and NH$_3$ that have only one.
Enhanced metallicity also creates a more oxidized gas, because we have assumed a solar C/O ratio less than unity.
These two trends work together to favor CO and especially CO$_2$ at higher metallicities, but
the two trends work to cross purposes with respect to hydrocarbon synthesis and soot formation,
because the more oxidized conditions suppress hydrocarbon formation.

\subsection{Hydrocarbons}

Methane polymerization begins with R45, ${\rm H} + {\rm CH}_4 \rightarrow {\rm H}_2 + {\rm CH}_3$.
Detailed analysis reveals that the chief pathway for making ${\rm C}_2{\rm H}_n$ hydrocarbons is the reaction between CH$_3$ radicals to make ethane,
$${\rm CH}_3 + {\rm CH}_3 + {\rm M} \rightarrow {\rm C}_2{\rm H}_6 {\rm M} . \eqno{\rm{R}289} $$
This came as something of a surprise to us because ethane is not itself very abundant in our models.
Evidently the more abundant ${\rm C}_2{\rm H}_n$ hydrocarbons are formed by reactions such as ${\rm C}_2{\rm H}_n+{\rm H} \rightarrow {\rm H}_2 + {\rm C}_2{\rm H}_{n-1}$, 
which are also pushed to the right by the high disequilibrium abundance of H.

\begin{figure}[!htb] 
   \centering
\includegraphics[width=1.00\textwidth]{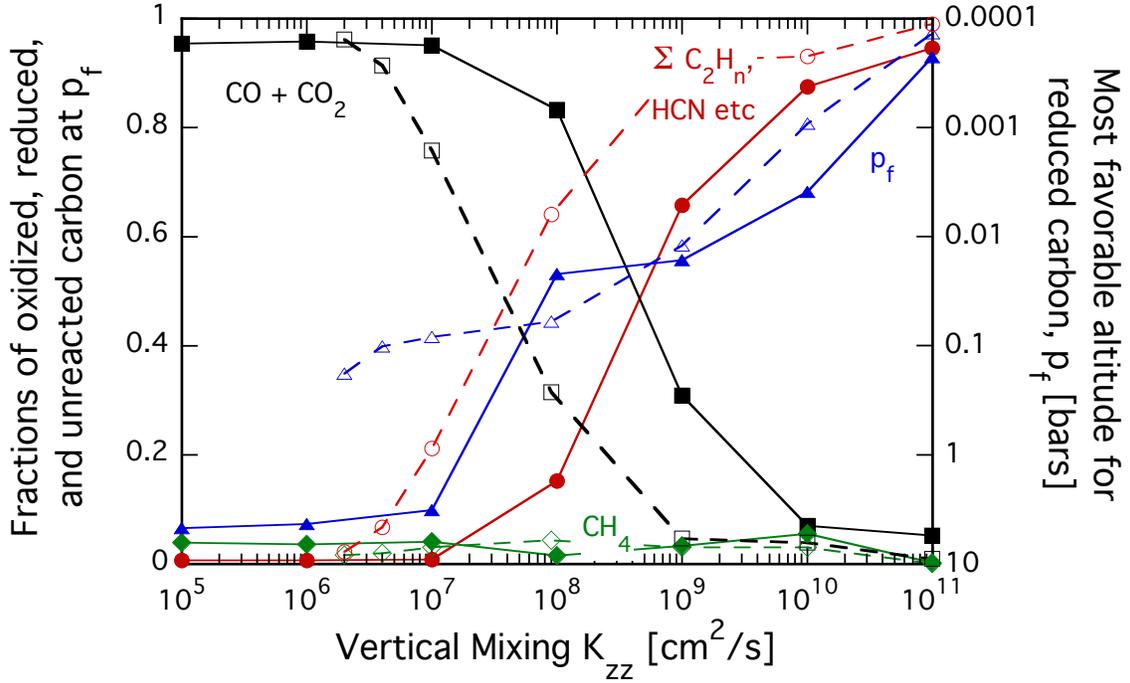} 
   \caption{Methane's fates at the most favorable altitude $p_f$ for soot formation as a function of vertical mixing $K_{zz}$.
  Fates are shown at $T=1000$ K (solid lines) and $T=900$ K (dashed lines).
  Fixed model parameters are $[{\rm M}/{\rm H}]=0.7$, $S_{uv}=100$, $p_0=100$ bars, and $g=20$ m/s$^2$.
  Irreversible oxidization ($x_{{\rm CO}}$) is denoted ``CO and CO$_2$'', interesting products ($x_{{\rm C}_2}$) are denoted
  ``$\Sigma {\rm C}_2{\rm H}_n, {\rm HCN~etc}$''.
   High $K_{zz}$ favors organic molecules.
 At 1000 K, soot formation seems reasonably likely for $K_{zz}\!>\!10^8$ cm$^2$/s. 
  The right hand axis shows that $p_f$ is a monotonically increasing function of $K_{zz}$.
 }
\label{fig:model_C}  
\end{figure}

The influence of $K_{zz}$ on carbon chemistry is viewed from a different angle in Figure \ref{fig:model_C}.
Here we show how carbon is allocated at the altitude that is most favorable to hydrocarbon production.
To do this we divide carbon into 3 categories.
The primary category is CH$_4$, the source molecule.  The other two are oxidized species containing CO bonds (CO, CO$_2$) 
and the more reduced products that do not contain CO bonds (C$_2$H$_2$, C$_2$H$_4$, HCN, CS$_2$, etc).
Call these $x_{{\rm CH}_4}$, $x_{{\rm CO}}$, and $x_{{\rm C}_2}$, respectively.
Total carbon is $\Sigma_{\rm C} = x_{{\rm CH}_4} + x_{{\rm CO}} + x_{{\rm C}_2}$.
We define the most favorable altitude $p_f$ for organic synthesis to be where $x_{{\rm C}_2}/\Sigma_{\rm C}$ reaches its maximum value.
Whilst every C-O bond is doomed to end in the irreversible production of CO,
 it is possible that much of the C in $x_{{\rm C}_2}$ can end in larger molecules such as PAHs or in the irreversible production of refractory soots.

\begin{figure}[!htb] 
   \centering
    \includegraphics[width=1.0\textwidth]{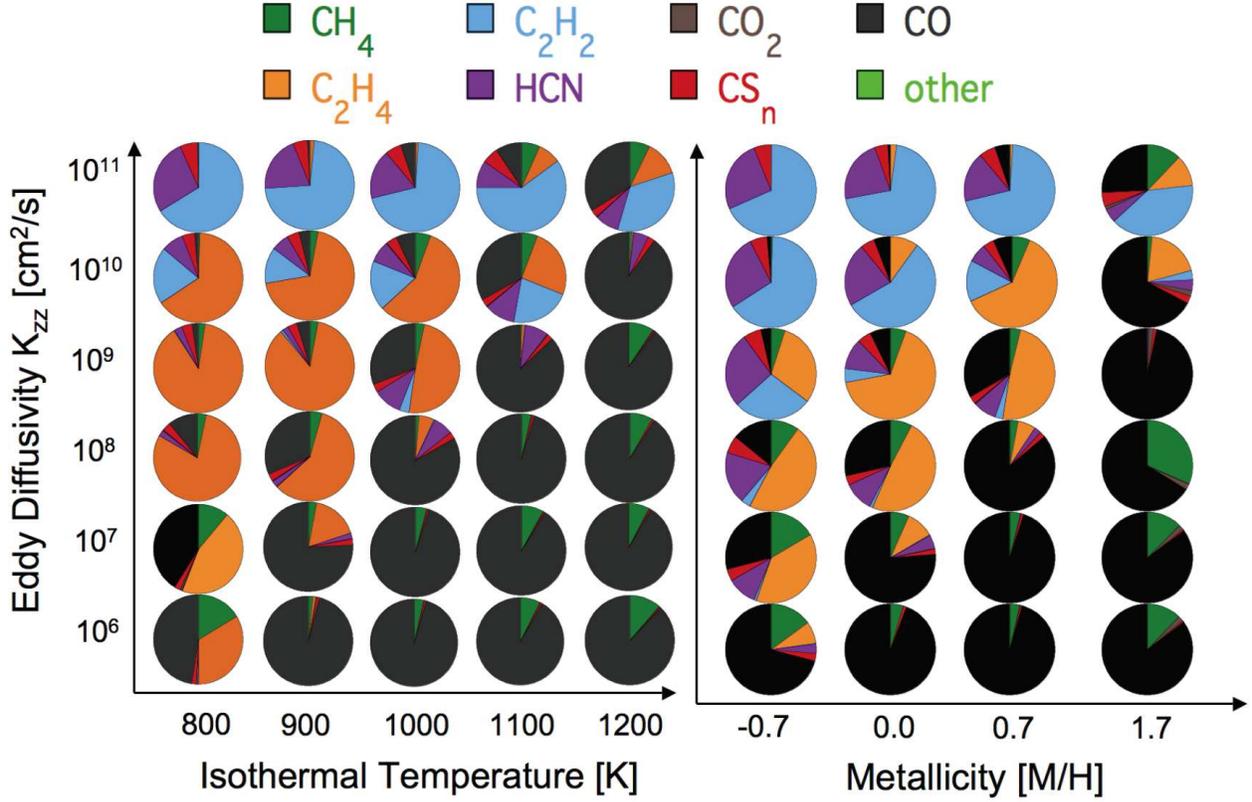}
   \caption{Generalization of Figure \ref{fig:model_C} to other temperatures ({\it left,} $[{\rm M}/{\rm H}]=0.7$) and metallicities ({\it right,} $T=1000$ K). 
   Common model parameters are $S_{uv}=100$, $p_0=100$ bars, $g=20$ m/s$^2$,
   and $\theta=30^{\circ}$ incidence angle for UV radiation.
   Each pie-chart shows how carbon is allocated at $p_f$ for a specific model.
   To first approximation higher values of $K_{zz}$ map to higher altitudes (lower $p_f$).
  {\it Left} Low temperatures and high $K_{zz}$ favor organic molecules.
   Very high values of $K_{zz} \geq 10^{10}$ cm$^2$/s mix CH$_4$ to altitudes that are more favorable to C$_2$H$_2$, the traditional soot precursor.  
{\it Right} Low metallicity favors hydrocarbons generally and C$_2$H$_2$ and HCN particularly. 
  High metallicity is quite oxidized.  The apparent preference for CH$_4$ at high metallicity and low K$_{zz}$ is an artifact of there being essentially no hydrocarbon synthesis and $p_f$ being near the lower boundary.}
\label{fig:figure_EM}  
\end{figure}

Figure \ref{fig:figure_EM} generalizes these results to different atmospheres as pie charts on $T$-$K_{zz}$ and $[{\rm M}/{\rm H}]$-$K_{zz}$ grids.
The relative amounts of carbon in different species is presented at the altitude $p_f$.
It is an optimistic figure, in the sense that it shows the most favorable conditions for carbon polymerization in each model.
Ethylene is favored by cooler atmospheres with strong vertical mixing, while CO and CO$_2$ are favored
 by higher temperatures, higher metallicity, and weaker vertical mixing.
 Acetylene --- soot's precursor --- is favored by very strong vertical mixing.  
The dependence of C$_2$H$_2$ itself on temperature is relatively weak, but the dependence of CO on temperature suggests
that competition with oxidation makes higher temperatures less suited to soot formation. 
Figure \ref{fig:figure_EM} also shows a clear negative dependence on metallicity
 that suggests that low metallicities are more favorable to soot formation.
The origin of the effect is that metals are on balance oxidizing because the C/O ratio in a solar gas is less than one.
Therefore adding more metals creates more oxidized conditions, other things equal.

Ethylene, acetylene, and hydrogen cyanide are first generation products of methane.
Benzene (C$_6$H$_6$) would be at best a third generation product (e.g., three acetylenes), and PAHs like coronene 
(C$_{24}$H$_{12}$) and ovalene (C$_{32}$H$_{14}$) 
are probably two or three generations further evolved.
At each step some carbon is lost to CO.
If it takes $n$ generations to make something resembling a soot, we might expect
the efficiency of soot formation to go like $x_{{\rm C}_2}^{n}$.  This is greater than 10\% for $x_{{\rm C}_2}>0.8$ and $n=10$,
so we infer that atmospheres where $x_{{\rm C}_2}>0.8$ have potential to develop significant organic hazes.    

PAHs interact strongly with visible light (Lou Allamandola, pers.\ comm.).
Coronene is golden yellow, ovalene is red, and still bigger ones tend to black.
Thus PAHs by themselves may suffice to explain high altitude optical absorption in hot Jupiters;
it may not be necessary that they agglomerate into grains, albeit grains are likely to be more refractory.

\begin{figure}[!htb]
 \centering
    \includegraphics[width=1.0\textwidth]{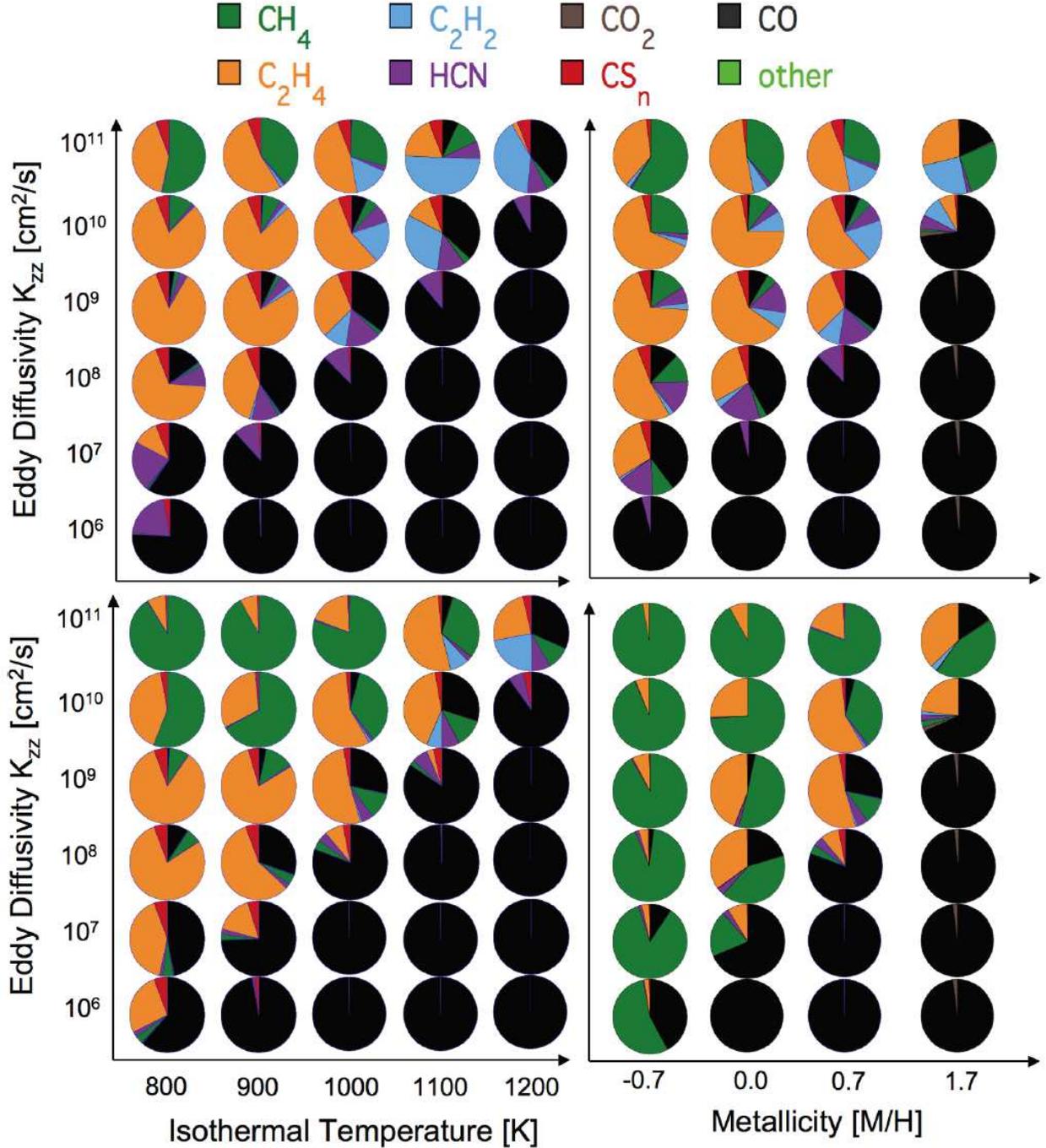}
  \caption{Carbon speciation at $\tau \approx 0.01$ (top) and $\tau \approx 1$ (bottom).
  These approximate what might be seen in grazing incidence in transit and in differential eclipses, respectively.
  For hydrocarbons vs. temperature, fixed parameters are $[{\rm M}/{\rm H}]=0.7$, $S_{uv}=100$, and $g=20$ m/s$^2$.
  For hydrocarbons vs. metallicity, fixed parameters are $T=1000$ K, $S_{uv}=100$, and $g=20$ m/s$^2$.   
}
\label{fig:figure_CC} 
\end{figure}

Figure \ref{fig:figure_CC} uses pie charts to provide an overview of the hydrocarbon chemistry that is much easier to interpret
than the spaghetti plots shown in Figures \ref{fig:figure_twelve}--\ref{fig:figure_Metals} (which are informative but daunting).
Here the pie charts show carbon speciation at two key altitudes, chosen to approximate the levels 
where the optical depths would be of order $0.01$ and $1$ in clear air.
Optical depths $\tau=0.01$ and $\tau=1$ suggest what might be seen in grazing incidence (primary eclipse) and in direct imaging (secondary eclipse), respectively.
 To estimate the optical depths we have approximated the pressure-dependent Rosseland infrared mean opacities from
Freedman et al (200x), for equilibrium chemistry for clear air of solar composition,
by $\kappa \approx 0.03 p^{0.6}$ cm$^{-2}$/g, where $p$ is in bars. 
This is a very crude fit, and the assumptions on which it is founded are of questionable relevance, 
 yet the results may have some value for illustrative purposes
 because the strongest infrared opacity source would in all cases be water vapor. 
Pressure levels corresponding to $\tau\approx 1$ and $\tau\approx 0.01$ for four different metallicities are listed in Table 4. 
Perhaps the most obvious and robust conclusion to draw from Figure \ref{fig:figure_CC} (and the analogous figures for nitrogen and sulfur discussed below) is that the apparent chemical composition of a planet depends on how it is viewed
(Fortney 2005). 
In Figure \ref{fig:figure_CC}, methane is much less prominent in transit observations than it is in observations of the photosphere.

\subsection{Nitrogen}

\begin{figure}[!htb]
 \centering
    \includegraphics[width=1.00\textwidth]{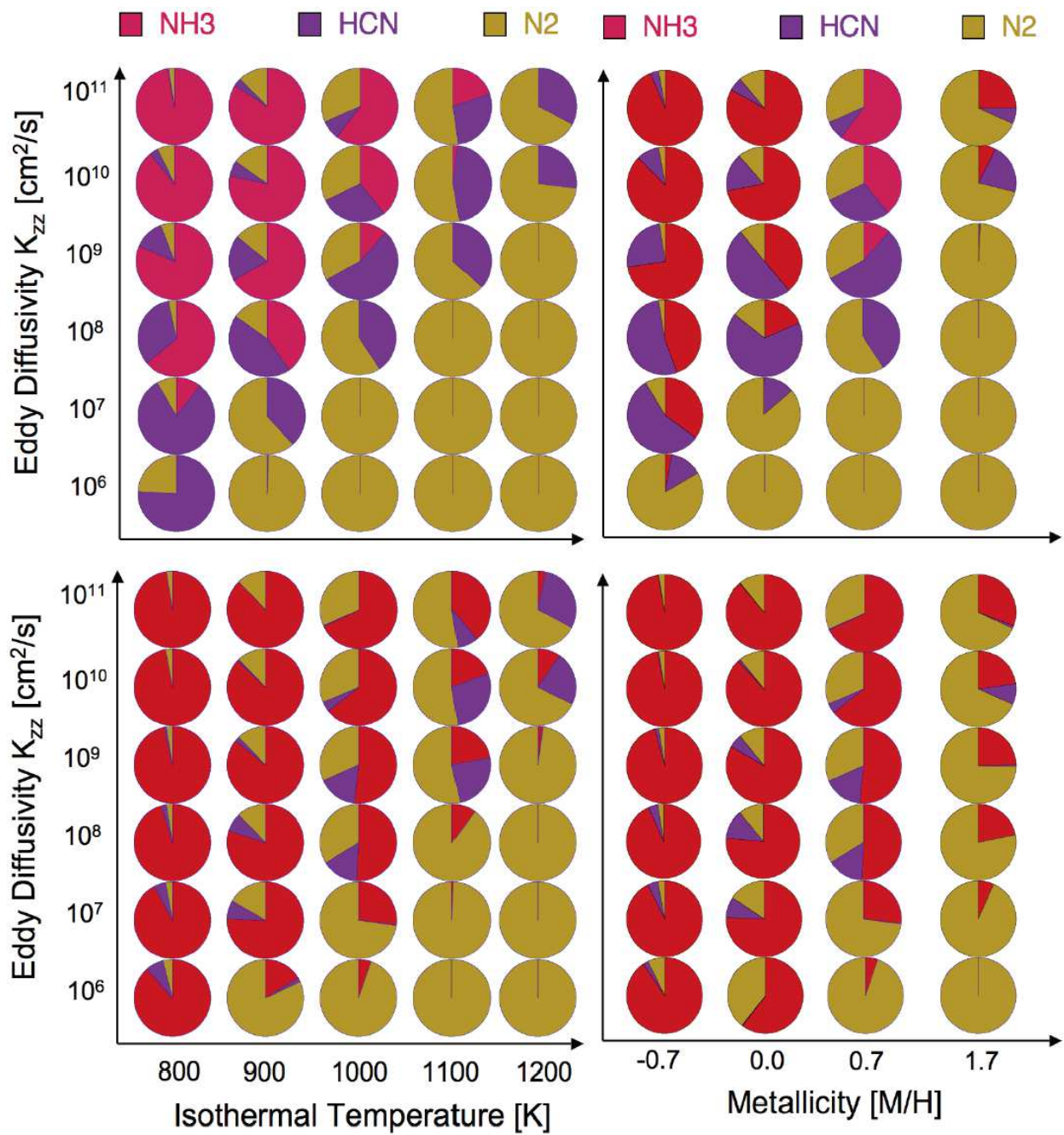} 
  \caption{Nitrogen speciation at $\tau \approx 0.01$ (top) and $\tau \approx 1$ (bottom).
  Cases and meanings are the same as for Figure \ref{fig:figure_CC}.  
  Ammonia is favored by low temperatures, low metallicity and strong vertical mixing, and N$_2$ is favored by the opposites.
  HCN, which forms most efficiently when ammonia is present and methane is polymerizing, is most abundant on the boundary between NH$_3$ and N$_2$.
  }
\label{fig:figure_NN} 
\end{figure}

 Figure \ref{fig:figure_NN} provides an overview of nitrogen speciation that is analogous to Figure \ref{fig:figure_CC} for carbon.
 Ammonia is favored by low temperatures, low metallicity and strong vertical mixing, and N$_2$ is favored by the opposites.
  HCN, which forms most efficiently when ammonia is present and methane is polymerizing, is most abundant on the boundary between NH$_3$ and N$_2$.
The chief pathway for making HCN in our models is 
$${\rm NH}_2 + {\rm C}_2{\rm H}_3 \rightarrow {\rm CH}_3 + {\rm HCN} . \eqno{\rm{R}337~} $$
This reaction is known to make an adduct; we have assumed that the C-N bond in the adduct leads inevitably to HCN,
much as we have assumed that the C-O bond in adducts made by OH and ${\rm C}_2{\rm H}_4$ or ${\rm C}_2{\rm H}_6$ 
lead inevitably to CO.  That a poorly characterized reaction like R337 should play a key role is not ideal.
Analogy to R289 suggests that the true path to HCN should pass through methylamine,
 ${\rm CH}_3 + {\rm NH}_2 \rightarrow {\rm CH}_3{\rm NH}_2$.
We have not yet included this pathway in our chemical scheme because we had no inkling that it should be important.
This means that our predictions for HCN abundances are more uncertain than those for the hydrocarbons,
and that our HCN abundances are likelier to be underestimated than overestimated.

\subsection{Sulfur}

\begin{figure}[!htb]
 \centering
    \includegraphics[width=1.0\textwidth]{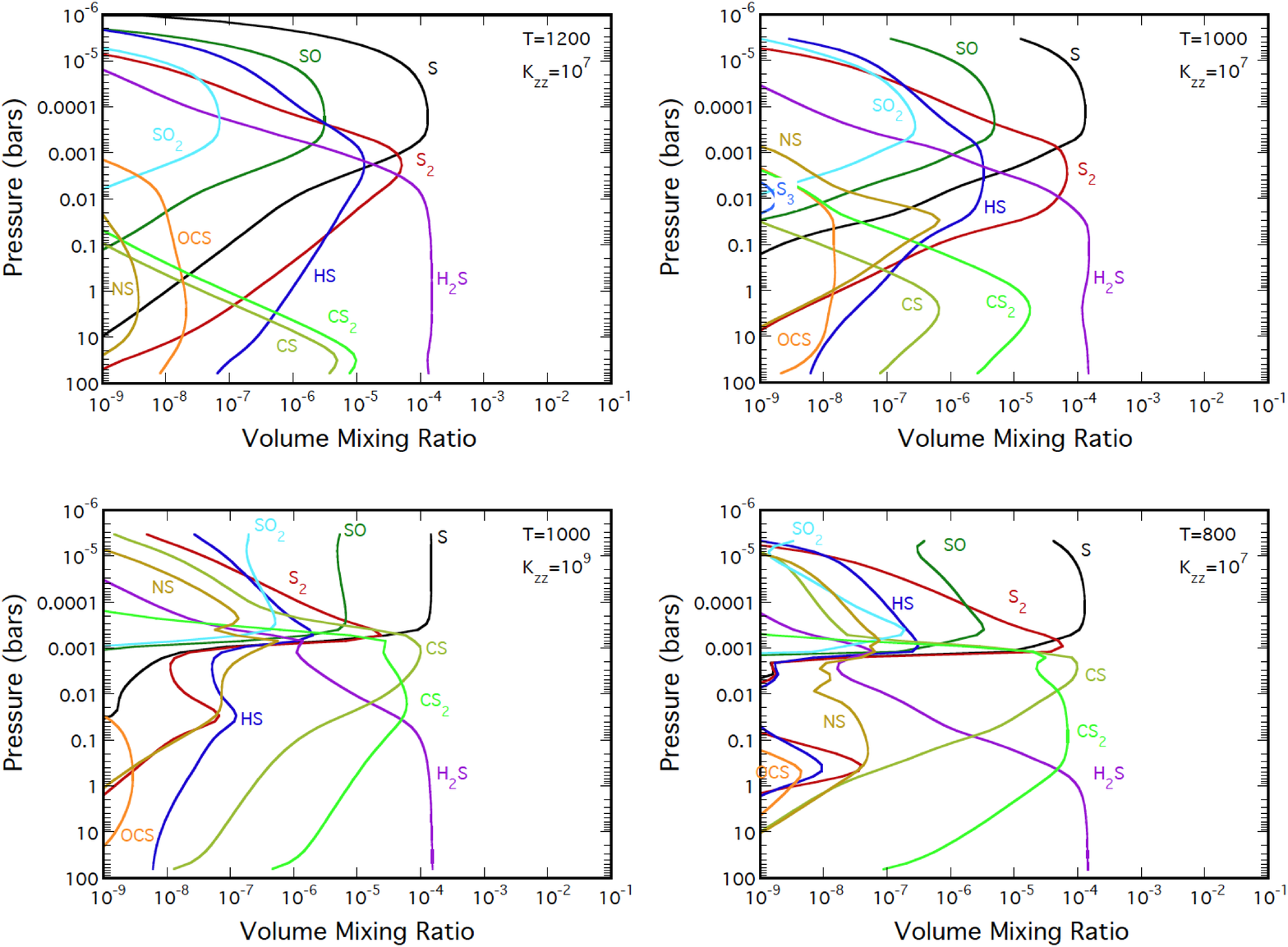} 
  \caption{Vertical profiles of sulfur species in isothermal atmospheres at different temperatures and vertical mixing.
  S$_2$ is favored by high temperatures and weak vertical mixing, while CS and CS$_2$ are favored
  by lower temperatures and strong vertical mixing. 
  H$_2$S is the stable gas in the deep atmosphere.  CS and CS$_2$ are found in environments that also favor C$_2$H$_4$.
  S$_2$ tends to form at the chemical boundary between the more reduced lower atmosphere and the more oxidized upper atmosphere.
}
\label{fig:figure_SK} 
\end{figure}

\begin{figure}[!htb]
 \centering
    \includegraphics[width=1.0\textwidth]{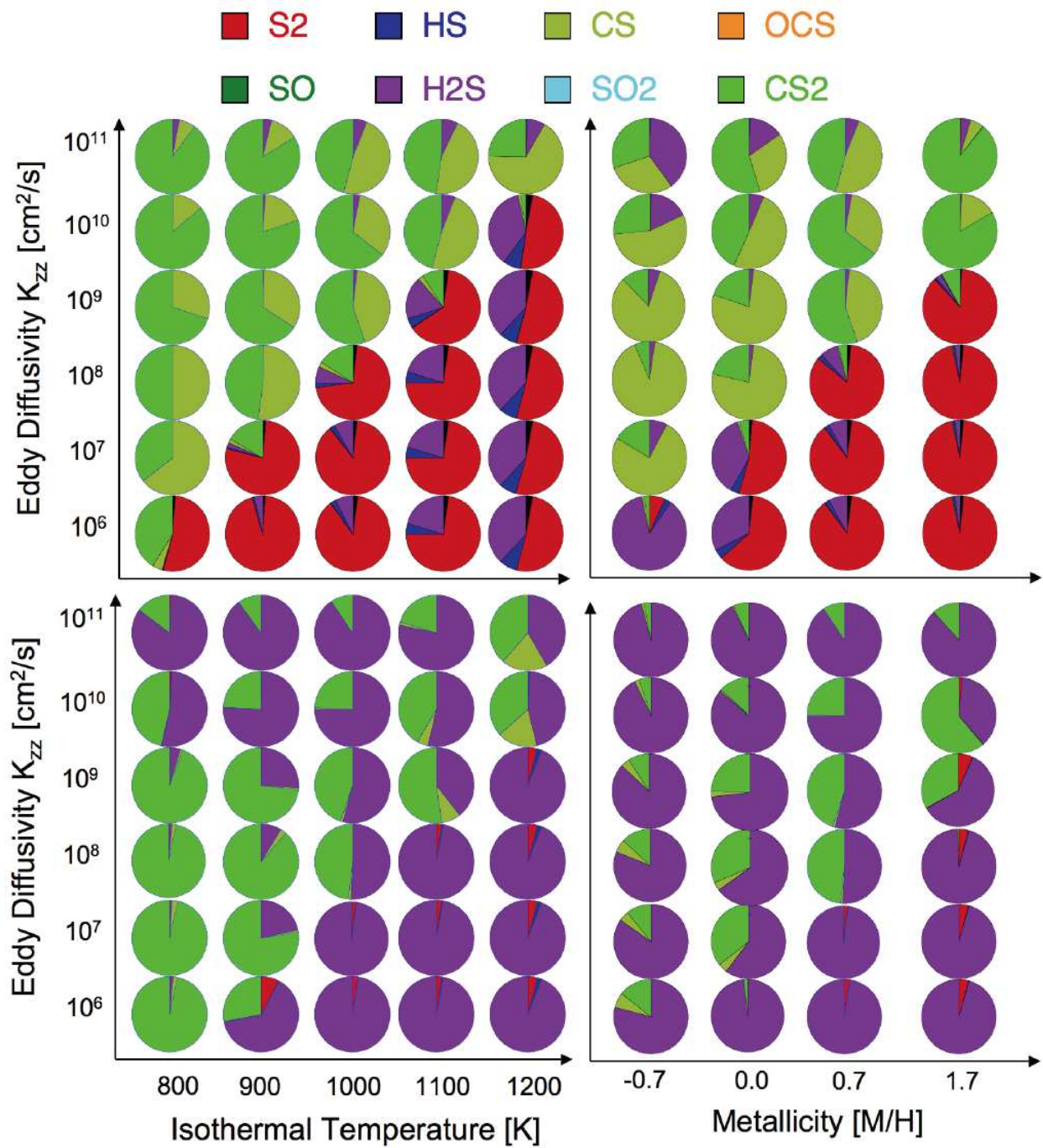}
  \caption{Sulfur speciation at $\tau \approx 0.01$ (top) and $\tau \approx 1$ (bottom).
  Cases and meanings are the same as for Figure \ref{fig:figure_CC}.  
  CS and CS$_2$ are favored by cooler atmospheres with strong vertical mixing, while S$_2$ is favored
  at high altitudes by higher temperatures, higher metallicities, and weaker vertical mixing. 
}
\label{fig:figure_SS} 
\end{figure}

Sulfur species are useful because they interact strongly with blue, violet, and UV light, which makes them potentially observable, especially in transit.
Sulfur speciation is illustrated in Figures \ref{fig:figure_SK} and \ref{fig:figure_SS}.
The former uses spaghetti plots to show the vertical structure of the sulfur species in a few
selected models.
The latter is an overview chart analogous to Figures \ref{fig:figure_CC} for carbon and \ref{fig:figure_NN} for nitrogen.
Equilibrium calculations predict that H$_2$S would be the most abundant S-containing species under nearly any condition encountered
in these atmospheres, as is seen above in Figure \ref{fig:figure_K}.

Zahnle et al (2009) discussed sulfur chemistry at higher temperatures ($T>1200$ K).
At these temperatures S$_2$ and HS are important at high altitudes but neither CS nor CS$_2$ were important.
In the cooler atmospheres investigated here both CS and CS$_2$ become important in parallel to hydrocarbons like ethylene, and for the same reasons:
the increased availability of reactive methane, and the decreased availability of OH from water.
Consequently cooler atmospheres with strong vertical mixing favor disequilibrium production of CS and CS$_2$.
Diatomic sulfur (S$_2$) is predicted to be most abundant at high altitudes at higher temperatures, higher metallicities, and weaker vertical mixing.
These dependencies are obvious in Figure \ref{fig:figure_SS}.
It is interesting that S$_2$ and acetylene (soot) appear to be mutually exclusive, the former indicating relatively metal-rich placid conditions, the latter indicating strong vertical mixing and low metallicity.

\section{Discussion}

The chemistry that we have modeled is unremarkable.
Most of the important reactions involving small organic molecules and radicals have been studied both experimentally and theoretically.
Thus the prediction that methane can react to make ethylene rather than CO seems secure.
The presumption that this will lead to bigger hydrocarbons than ethylene seems equally secure.
Whether this process continues up to PAHs and soot is less secure.

On the surface our results appear to disagree with what Liang et al (2004) found previously.
However, there are enough differences between their simulations and ours that the apparent disagreement is misleading.
The higher hydrocarbons that form in our models are primarily products of disequilibrium thermochemistry, not photochemistry.
They form because recombination of H to H$_2$ is kinetically inhibited and consequently the atomic hydrogen density
 is much higher than it would be in equilibrium.
 Liang et al find less H than we do because they did not include sulfur.
 The most important reactions governing the H abundance are those involving sulfur, especially
$${\rm H} + {\rm H}_2{\rm S} \leftrightarrow {\rm H}_2 + {\rm HS} \eqno{\rm{R}156} $$
$${\rm H}_2 + {\rm HS} \leftrightarrow {\rm H}_2{\rm S} + {\rm H} \eqno{\rm{R}157} $$
and
$${\rm S} + {\rm H}_2 \leftrightarrow {\rm H} + {\rm HS} \eqno{\rm{R}168} $$
$${\rm H} + {\rm HS} \leftrightarrow {\rm H}_2 + {\rm S} . \eqno{\rm{R}169} $$
Reactions R156 and R157 are generally the most frequently occurring reactions in the atmosphere by a wide margin (Zahnle et al 2009). 
Reactions R168 and R169 are most important at higher altitudes (lower pressures).
Photolysis has little to do with these.

Another difference between our study and Liang et al (2003, 2004) is that Liang et al
 attempt to use realistic approximations to $T(z)$ and $K_{zz}(z)$.
Constant temperature and constant $K_{zz}$ is a feature of our study.  
Our emphasis here is on the survey of parameter space.
Temperature and vertical mixing are the two key parameters.
The cleanest way to look at the problem is therefore to use $T$ and $K_{zz}$ as the independent variables.
More complicated atmospheric profiles of either $T(z)$ or $K_{zz}(z)$ requires introducing additional parameters,
which introduces an additional level of complexity that may not be justified.

For $K_{zz}$, Liang et al use $K_{zz}\propto N^{-0.6}$.  Their assumption gives $K_{zz} = 2.4\times 10^7$ cm$^2$/s
at 1 bar, $1.5 \times 10^9$ cm$^2$/s at 1 mbar, and $1.0 \times 10^{11}$ cm$^2$/s at 1 $\mu$bar (all at 1200 K).
This range is similar to our high $K_{zz}$ cases.
For $T(z)$, Liang et al use three specific profiles derived from detailed radiative transfer model results.
All three are very hot at 1 bar (the coolest is 1800 K). 
These temperatures are much higher than anything we address here.  
In our model methane is immediately destroyed if $T>1200$ K.
We agree with Liang et al (2004) that very hot planets ought not to have hydrocarbon hazes.

As a test we computed chemistry using the nominal $p$-$T$ profile from Fortney et al's (2009) most recent model of HD 189733b.  Their model is hot ($T\approx 1500$ K) below 3 bars and cool ($T\approx 900$ K) above 10 mbars, with a smooth transition in between. 
We note that Fortney et al include only Rayleigh scattering, so that their computed albedo is relatively low,
and thus their $T(z)$ may not closely resemble that of a hazy planet.
For the simulation we set $K_{zz}=10^8$.  We find very little methane above the lower boundary and no significant polymerization.  We then reduced the lower temperature to 1200 K, and then to 1100 K.  At 1200 K, NH$_3$ is abundant but hydrocarbons are not, while at 1100 K we see rampant hydrocarbon polymerization. 
These results are quite consistent with Figure \ref{fig:figure_EM}. 

In passing, we note that solar C/O is also an assumption.  Snowline models of giant planet formation can give lower C/O, while the less well known tar-line model (Lodders 2004) predicts higher C/O.  
The true C/O ratio of Jupiter is not yet known.
High C/O changes the chemistry dramatically; among other things, soots would be strongly favored at higher temperatures.  The possibility that a planet like HD 189733b might have high C/O should be in play.        

\section{Conclusions}

We find that vertical mixing in hot Jupiters with temperatures below 1000 K 
can provide environments suitable for turning methane
into ethylene, acetylene, and hydrogen cyanide.  
We speculate that such atmospheres would be conducive to generating PAHs and perhaps soots.
We also reiterate the point that planetary compositions inferred from transit observations,
 which probe high altitudes (low pressures), 
 can differ markedly from those inferred from reflected or emitted light from the same planet.
In general, thermal chemistry and vertical mixing are more important to disequilibrium chemistry in warm or hot Jupiters than is photochemistry.
 Photochemistry plays a bigger role at low temperatures.

The presence of abundant PAHs or soots at high altitudes could account for the up-to-now unique properties of HD 189733b, if HD 189733b proves to be cool enough, or its metallicity low enough, or its C/O ratio high enough, or its vertical mixing as vigorous as Showman et al (2009) suggest.
This would make HD 189733b the first of a new class of warm Jupiters.
As noted, Swain et al (2009) reported that HD 189733b's metallicity might be rather low.
On the other hand HD 189733 is a relatively strong source of UV (Knutson 2010),
which makes photolysis rates high, a factor that we have shown will suppress high altitude soot formation.
Whether our thermometric interpretation of HD 189733b actually applies to HD 189733b (recent models suggest that the planet is too hot if it has solar C/O) is not the main point:  many organic-rich hazy
warm Jupiters are sure to be discovered in the near future.

\section{Acknowledgements}
KJZ thanks NASA's Exobiology Program for support.
MSM thanks NASA's Planetary Atmospheres Program for support

\newpage

\linespread{1.0}

\begin{table}[htdp]
\begin{tabular}{ll} 
\multicolumn{2}{c}{\bf Table 1}\\
\multicolumn{2}{c}{Chemical Species}\\
\hline
\hline
Elements &  Species\\
\hline
H, O & H, H$_2$, O, OH, H$_2$O, O$_2$, O($^1$D) \\
CO  & CO, CO$_2$, HCO, H$_2$CO, H$_2$COH, CH$_3$O, CH$_3$OH \\
CH$_n$  & C, CH, CH$_2$, $^1$CH$_2$, CH$_3$, CH$_4$ \\
C$_m$H$_n$  & C$_2$, C$_2$H, C$_2$H$_2$, C$_2$H$_3$, C$_2$H$_4$, C$_2$H$_5$, C$_2$H$_6$, C$_2$H$_2$OH, C$_2$H$_4$OH, C$_4$H, C$_4$H$_2$\\
N  & N, N$_2$, NH, NH$_2$, NH$_3$, CN, HCN, H$_2$CN, HCNOH, NO, NS \\
S  & S, HS, H$_2$S, S$_2$, S$_3$, S$_4$, S$_8^a$, S$_8^{\ast b}$, CS, CS$_2$, OCS, HCS, H$_2$CS, SO, HSO, SO$_2$\\
\hline
\multicolumn{2}{l}{$a$ -- Ring (ground state)}\\
\multicolumn{2}{l}{$b$ -- Linear (excited state)}\\
\end{tabular}

\end{table}

\begin{table}[htdp]  
\begin{tabular}{l lcl l p{1.5cm} } 
\multicolumn{6}{c}{\bf Table 2}\\
\multicolumn{6}{c}{Chemical Reactions}\\
\hline
\hline
   & Reactants$^a$  &  & Products & Rate$^b$ & Reference \\
\hline
 R1   & H            + H            + M & $\!\!\!\rightarrow$ &  H$_2$        + M &$  8.8\!\times\! 10^{-33} \left(T/298 \right)^{-0.60}$ & Ba92\\
        & H            + H           &$\!\!\!\rightarrow$&  H$_2$        + M &$  2.0\!\times\! 10^{-10}$ & \\
 R2   & H$_2$        + M           &$\!\!\!\rightarrow$ &  H            + H                          +M           & $  1.5\!\times\! 10^{-09} e^{-48400/T}$ & Ba92\\
 R3   & H$_2$        + O           & $\!\!\!\rightarrow$ &  OH           + H                                       & $  3.5\!\times\! 10^{-13} \left(T/298\right)^{ 2.67}e^{ -3160/T}$ & Ba92\\
 R4   & H            + OH          & $\!\!\!\rightarrow$ &  H$_2$        + O                                       & $  1.7\!\times\! 10^{-13} \left(T/298\right)^{ 2.64}e^{ -2240/T}$ & rev3\\
 R5   & H$_2$        + OH          & $\!\!\!\rightarrow$ &  H$_2$O       + H                                       & $  1.6\!\times\! 10^{-12} \left(T/298\right)^{ 1.60}e^{ -1660/T}$ & Ba92\\
 R6   & H$_2$O       + H           & $\!\!\!\rightarrow$ &  H$_2$        + OH                                      & $  6.9\!\times\! 10^{-12} \left(T/298\right)^{ 1.60}e^{ -9270/T}$ & Ba92\\
 R7   & OH           + OH          & $\!\!\!\rightarrow$ &  H$_2$O       + O                                       & $  1.7\!\times\! 10^{-12} \left(T/298\right)^{ 1.14}e^{   -50/T}$ & Ba92\\
 R8   & H$_2$O       + O           & $\!\!\!\rightarrow$ &  OH           + OH                                      & $  1.8\!\times\! 10^{-11} \left(T/298\right)^{ 0.95}e^{ -8570/T}$ & Li91\\
 R9   & O            + H            + M & $\!\!\!\rightarrow$ &  OH           + M &$  4.3\!\times\! 10^{-32} \left(T/298 \right)^{-1.00}$ & Ts86\\
         & O            + H           &$\!\!\!\rightarrow$&  OH           + M &$  2.0\!\times\! 10^{-11}$ & \\
 R10   & OH           + M           &$\!\!\!\rightarrow$ &  O            + H                          +M           & $  4.0\!\times\! 10^{-09} e^{-50000/T}$ & Ts86\\
 R11   & H            + O           & $\!\!\!\rightarrow$ &  OH           + h$\nu$                                     & $  9.9\!\times\! 10^{-19} \left(T/298 \right)^{-0.38}$ & Mi97\\
 R12   & O            + O            + M & $\!\!\!\rightarrow$ &  O$_2$        + M &$  1.0\!\times\! 10^{-33} \left(T/298 \right)^{-1.00}$ & Wa84\\
           & O            + O           &$\!\!\!\rightarrow$&  O$_2$        + M &$  2.0\!\times\! 10^{-11}$ & \\
 R13   & O$_2$        + M           & $\!\!\!\rightarrow$ &  O            + O                          +M           & $  1.0\!\times\! 10^{-08} \left(T/298\right)^{-1.00}e^{-59400/T}$ & Ts86\\
 R14   & O            + OH          &$\!\!\!\rightarrow$ &  O$_2$        + H                                       & $  2.4\!\times\! 10^{-11} e^{  -353/T}$ & Ba92\\
 R15   & H            + O$_2$       &$\!\!\!\rightarrow$ &  OH           + O                                       & $  3.3\!\times\! 10^{-10} e^{ -8460/T}$ & Ba92\\
 R16   & H$_2$        + O$_2$       & $\!\!\!\rightarrow$ &  OH           + OH                                      & $  4.0\!\times\! 10^{-11} \left(T/298\right)^{ 0.47}e^{-35100/T}$ & Ka05\\
 R17   & CO           + M           & $\!\!\!\rightarrow$ &  C            + O                          +M           & $  2.7\!\times\! 10^{-03} \left(T/298\right)^{-3.52}e^{130000/T}$ & Ba92\\
 R18   & C            + O           &$\!\!\!\rightarrow$&  CO           + M &$  0 $ & \\
           & C            + O           &$\!\!\!\rightarrow$&  CO           + M &$  0 $ & \\
 R19   & C            + OH          & $\!\!\!\rightarrow$ &  CO           + H                                       & $  1.1\!\times\! 10^{-10} \left(T/298 \right)^{ 0.50}$ & Mi97\\
 R20   & C            + OH          & $\!\!\!\rightarrow$ &  CH           + O                                       & $  2.2\!\times\! 10^{-11} \left(T/298\right)^{ 0.50}e^{-14800/T}$ & Mi97\\
 R21  & C            + O$_2$       &$\!\!\!\rightarrow$ &  CO           + O                                       & $  1.6\!\times\! 10^{-11}$ & Ba92\\
 R22   & H            + CO          & $\!\!\!\rightarrow$ &  C            + OH                                      & $  2.8\!\times\! 10^{-09} \left(T/298\right)^{ 0.50}e^{-76900/T}$ & rev21\\
 R23   & CO           + OH          & $\!\!\!\rightarrow$ &  CO$_2$       + H                                       & $  1.8\!\times\! 10^{-14} \left(T/298\right)^{ 1.89}e^{   583/T}$ & Li07\\
 R24   & CO$_2$       + H           & $\!\!\!\rightarrow$ &  CO           + OH                                      & $  6.1\!\times\! 10^{-11} \left(T/298\right)^{ 0.64}e^{-12500/T}$ & rev23\\
 \ldots   & \ldots       & \ldots &  \ldots                                      & \ldots & \ldots \\
 \ldots   & \ldots       & \ldots &  \ldots                                      & \ldots & \ldots \\
 \ldots   & \ldots       & \ldots &  \ldots                                      & \ldots & \ldots \\
 R527  & C$_2$H$_5$          + H$_2$CN    &$\!\!\!\rightarrow$ &  HCN    +  C$_2$H$_6$                               & $ 7.7\!\times\! 10^{-12} $ &  \\
 R528  & NH$_2$      + C$_2$H$_5$    &$\!\!\!\rightarrow$ &  NH    +  C$_2$H$_6$                 & $ 3.0\!\times\! 10^{-13} \left(T/298 \right)^{1.00}e^{  -4400/T}$ & Xu99\\
\hline
\multicolumn{6}{l}{ $a$ --- M refers to the background atmosphere, principally H$_2$ and He; units of density [cm$^{-3}$].}\\
\multicolumn{6}{l}{$b$ --- 2-body reaction rates are in cm$^{3}$s$^{-1}$;  3-body rates are in cm$^{6}$s$^{-1}$.}\\
\end{tabular}

\end{table}

\begin{table} 
\begin{tabular}{l lcl l p{1.5cm} } 
\multicolumn{6}{c}{\bf Table 3}\\
\multicolumn{6}{c}{Photolysis Reactions}\\
\hline
\hline
 & Species  &  & Products & Rate$^a$ & Reference \\
\hline 
 P1  & H$_2$O       + h$\nu$         &$\!\!\!\rightarrow$ &  OH           + H                                       & $  1.6\!\times\! 10^{-03}$ & Sa03\\ %
 P2  & CO$_2$       + h$\nu$         &$\!\!\!\rightarrow$ &  CO           + O                                       & $  2.7\!\times\! 10^{-07}$ & Ok78,Hu92\\ %
 P3  & CO$_2$       + h$\nu$         &$\!\!\!\rightarrow$ &  CO           + O($^1$D)                                  & $  3.0\!\times\! 10^{-05}$ & Ok78,Hu92\\ %
 P4  & O$_2$        + h$\nu$         &$\!\!\!\rightarrow$ &  O            + O                                       & $  9.6\!\times\! 10^{-06}$ & Sa03\\ %
 P5  & O$_2$        + h$\nu$         &$\!\!\!\rightarrow$ &  O            + O($^1$D)                                  & $  5.1\!\times\! 10^{-04}$ & Sa03\\ %
 P6  & NO           + h$\nu$         &$\!\!\!\rightarrow$ &  N            + O                                       & $  3.7\!\times\! 10^{-06}$ & Sa03 \\ %
 P7  & H$_2$S       + h$\nu$         &$\!\!\!\rightarrow$ &  HS           + H                                       & $  2.7\!\times\! 10^{-02}$ & Ok78,Hu92\\ %
 P8  & NH$_3$       + h$\nu$         &$\!\!\!\rightarrow$ &  NH$_2$       + H                                       & $  1.1\!\times\! 10^{-02}$ & Ok78,Hu92\\ %
 P9  & NH$_3$       + h$\nu$         &$\!\!\!\rightarrow$ &  NH           + H$_2$                                   & $  5.5\!\times\! 10^{-03}$ & Ok78,Hu92\\ %
 P10  & CH$_4$       + h$\nu$         &$\!\!\!\rightarrow$ &  CH$_2$       + H$_2$                                   & $  3.9\!\times\! 10^{-04}$ & Hu92\\ %
 P11  & CH$_4$       + h$\nu$         &$\!\!\!\rightarrow$ &  CH$_3$       + H                                       & $  3.9\!\times\! 10^{-04}$ & Hu92\\ %
 P12  & SO$_2$       + h$\nu$         &$\!\!\!\rightarrow$ &  SO           + O                                       & $  1.7\!\times\! 10^{-02}$ &  Ok78,Hu92\\ %
 P13  & SO$_2$       + h$\nu$         &$\!\!\!\rightarrow$ &  S            + O$_2$                                   & $  6.9\!\times\! 10^{-04}$ &  Ok78,Hu92\\ %
 P14  & SO$_2$       + h$\nu$         &$\!\!\!\rightarrow$ &  S            + O                          +O           & $  1.7\!\times\! 10^{-05}$ &  Ok78,Hu92\\ %
 P15  & SO           + h$\nu$         &$\!\!\!\rightarrow$ &  S            + O                                       & $  4.4\!\times\! 10^{-02}$ & Ok78,Hu92\\ %
 P16  & CS$_2$       + h$\nu$         &$\!\!\!\rightarrow$ &  CS           + S                                       & $  3.9\!\times\! 10^{-01}$ & Mo81,Ah92\\ %
 P17  & OCS          + h$\nu$         &$\!\!\!\rightarrow$ &  CO           + S                                       & $  2.5\!\times\! 10^{-03}$ & Sa03\\ %
 P18  & S$_2$        + h$\nu$         &$\!\!\!\rightarrow$ &  S            + S                                       & $  1.7\!\times\! 10^{-01}$ & Za09\\ %
 P19  & S$_3$        + h$\nu$         &$\!\!\!\rightarrow$ &  S$_2$        + S                                       & $  1.1\!\times\! 10^{+02}$ & Za09\\ %
 P20  & S$_4$        + h$\nu$         &$\!\!\!\rightarrow$ &  S$_3$        + S                                       & $  1.1\!\times\! 10^{+01}$ & Za09\\ %
 P21  & S$_8$        + h$\nu$         &$\!\!\!\rightarrow$ &  S$_8^{\ast}$                                         & $  1.7\!\times\! 10^{+00}$ & Ka89\\ %
 P22  & S$_8^{\ast}$       + h$\nu$         &$\!\!\!\rightarrow$ &  S$_4$        + S$_4$                                   & $  3.5\!\times\! 10^{+00}$ & Ka89\\ %
 P23  & C$_2$H$_2$   + h$\nu$         &$\!\!\!\rightarrow$ &  C$_2$H       + H                                       & $  2.4\!\times\! 10^{-03}$ & Ok78,Hu92\\ %
 P24  & C$_2$H$_4$   + h$\nu$         &$\!\!\!\rightarrow$ &  C$_2$H$_3$   + H                                       & $  5.3\!\times\! 10^{-03}$ & Ok78,Hu92\\ %
 P25  & C$_2$H$_6$   + h$\nu$         &$\!\!\!\rightarrow$ &  C$_2$H$_5$   + H                                       & $  1.1\!\times\! 10^{-03}$ & Ok78,Hu92\\ %
 P26  & C$_4$H$_2$   + h$\nu$         &$\!\!\!\rightarrow$ &  C$_4$H       + H                                       & $  4.8\!\times\! 10^{-03}$ & note \\ %
 P27  & H$_2$CO      + h$\nu$         &$\!\!\!\rightarrow$ &  CO           + H$_2$                                   & $  6.1\!\times\! 10^{-03}$ & Sa03\\ %
 P28  & H$_2$CO      + h$\nu$         &$\!\!\!\rightarrow$ &  HCO          + H                                       & $  7.1\!\times\! 10^{-03}$ & Sa03\\ %
 P29  & H$_2$CO      + h$\nu$         &$\!\!\!\rightarrow$ &  CO           + H                          +H           & $  0$ & Sa03 \\ %
 P30  & HCN          + h$\nu$         &$\!\!\!\rightarrow$ &  CN           + H                                       & $  2.5\!\times\! 10^{-03}$ & Hu92\\ %
 P31  & HSO          + h$\nu$         &$\!\!\!\rightarrow$ &  HS           + O                                       & $ 6.1\!\times\! 10^{-02}$ & note \\ %
 P32  & HS           + h$\nu$         &$\!\!\!\rightarrow$ &  H            + S                                       & $  1.0\!\times\! 10^{+01}$ & Za09\\ %
 P33  & CH$_4$       + h$\nu$         &$\!\!\!\rightarrow$ &  CH           + H$_2$                      +H           & $  1.9\!\times\! 10^{-04}$ & Ok78,Hu92\\ %
\hline
\multicolumn{6}{l}{$a$  Photolysis rates at the top of the atmosphere for $S_{uv}=100$ and $30^{\circ}$ zenith angle. }\\
\multicolumn{6}{l}{ P21. Photolysis of S$_8$ is presumed to create the linear radical S$_8^{\ast}$.  This is a placeholder at high}\\
\multicolumn{6}{l}{~temperatures where S$_8$ is not expected and at low temperatures where S$_8$ condenses.}\\
\multicolumn{6}{l}{ P26. Assumes twice the cross section of C$_2$H$_2$.}\\
\multicolumn{6}{l}{ P31. Assumes cross section of HO$_2$.}\\
\end{tabular}
\end{table}%

\begin{table}[htdp]
\begin{tabular}{ccc} 
\multicolumn{3}{c}{\bf Table 4}\\
\multicolumn{3}{c}{Reference Pressures}\\
\hline
\hline
   & $\tau=0.01$ &   $\tau=1$\\
\hline
 $[{\rm M}/{\rm H}]$ & $p$ [mbars] &  $p$ [mbars]\\
\hline
0.2 & 30 & 500 \\
1  & 10 & 200 \\
5  & 4 & 70 \\
50  & 1 & 16 \\
\end{tabular}

\end{table}

\end{document}